**Effect of TCNQ layer cover on oxidation dynamics of black phosphorus**

*Laura Meingast, Maria Koleśnik-Gray, Martin Siebert, Gonzalo Abellán, Stefan Wild, Vicent Lloret, Udo Mundloch, Frank Hauke, Andreas Hirsch and Vojislav Krstić\**

L. Meingast, Dr. M. Koleśnik-Gray, M. Siebert, Prof. V. Krstić
Department of Physics, Friedrich-Alexander-Universität Erlangen-Nürnberg (FAU), Staudtstr. 7, 91058 Erlangen, Germany
E-mail: vojislav.krstic@fau.de
Dr. Gonzalo Abellan, S. Wild, V. Lloret, U. Mundloch, Dr. F. Hauke, Prof. A. Hirsch
Department of Chemistry and Pharmacy and Joint Institute of Advanced Materials and Processes (ZMP), Friedrich-Alexander University of Erlangen-Nürnberg (FAU), Henkestr. 42, 91054 Erlangen, Germany

Keywords: black phosphorus, TCNQ, oxidation, dynamics

Abstract

The puckered surface of black phosphorus represents an ideal substrate for an unconventional arrangement of physisorbed species and the resulting specific two-dimensional chemistry of this system. This opens the way to investigate the chemical and physical properties of locally confined areas of black phosphorus without the necessity for further physical downscaling of the material. We have evaporated TCNQ on top of black phosphorus under over-saturation non-equilibrium conditions in vacuum. The evolution of linear density and height of droplets formed through oxidation during exposure to air was studied time-dependently by scanning-force microscopy. Our study suggests that the TCNQ molecules spontaneously arrange in a thin layer of the order of a few nm height, which, however, is fragmented with a periodicity of about 100 nm. It is shown that within the confined space separating the layer fragments the chemical dynamics of the oxidation process is remarkably different than on a bare black phosphorus surface.

Black phosphorus (BP) has attracted significant scientific interest as member of the family of so-called van-der-Waals two-dimensional (2D) materials constituting of only one element [1-9]. The electronic properties of thin, few-layer BP and its monolayer phosphorene have been in the main focus of research[10-19]. The uniqueness of the puckered BP surface which in principle represents a one-dimensional periodic potential modulation[20-22] has only recently been shifting into focus regarding non-covalent functionalization with molecules in the monolayer limit [23,24] in contrast to graphene [25-28]. Such an electrostatically pre-structured surface offers a large variety of opportunities for studying the effects of confined surface-chemical reactions, electrochemistry as well as optoelectronic properties of the surface. Therefore, the natural pre-structuring of the BP surface could be the source of unconventional arrangements of molecules on it, not only in the monolayer limit, but also for molecular layers of nm height. This arrangement principally can span from ordering along the potential directions indicated on the surface to the formation of layered structures with confined uncovered areas where the BP surface is still exposed to the environment and thus freely accessible. While the first type of arrangement is what would be typically expected within a one-dimensional potential lattice, the fragmented layer-like arrangement is more appealing: It provides the possibility to define (comparably larger) confined areas of BP without a structural edge. These areas are qualitatively different from *e.g.* BP flakes mechanically broken down into smaller pieces as the surface of the BP extends beneath the layer and therefore are both spatially and electrostatically confined (*cf.* localized interaction/charge-transfer between BP and molecules[29]). Therefore, the electronic, optical and chemical properties of such confined areas are expected to be different and can potentially open new ways towards applications involving BP.

Targeting the formation of fragmented-layer-like arrangement on BP, we chose tetracyanoquinodimethane (TCNQ) and sublimated it under vacuum conditions on top of BP. The change in chemical reactivity of the areas between TCNQ-layer fragments could be directly traced by detecting the time-resolved formation of phosphoric and phosphorous acids/water droplets as a consequence of the dissociative chemisorption of $O_2$ on BP surface under ambient conditions [30-34] using scanning force microscopy. Our study indicates that indeed spontaneously a large-scale periodic fragmentation of the TCNQ layer occurs with characteristic fragment dimensions of the order of 100 nm on top of the BP surface. In particular, we also observed that the oxidation of the BP within the confined areas between fragments follows a significantly slower dynamics than in extended BP surfaces. We attributed this qualitative change in behavior to the difference in electrostatics within a confined area compared to an extended BP surface.

**Experimental Methods**

*BP exfoliation and TCNQ deposition*:

Deposition of TCNQ on BP samples was carried out at $10^{-3}$ mbar at a sublimation temperature of 90 °C for one hour in a glass flask. As substrate a standard $SiO_2$(300 nm)/Si wafer was used on which the BP flakes have been deposited by mechanical cleavage in Ar atmosphere prior to TCNQ exposure.

*Raman measurements and AFM analysis*:

The confirmation of the flakes to be BP and covered by TCNQ was obtained by Raman-spectroscopy (laser line 532 nm). The surface characterization of the BP with and without TCNQ was carried out by scanning force microscopy (AFM) to avoid additional degradation mechanisms as would be induced by laser in Raman microscopy. These measurements were carried out under ambient conditions to allow for monitoring of the oxidation process of the different surfaces. The resulting images were analyzed via Fourier-transformation using the standard AFM-tool software (open source [35]). We emphasize that we avoided the use of scanning Raman microscopy for monitoring the degradation with time as this could affect the oxidation behavior itself.

TCNQ is a favorable molecule because it has been theoretically[29] predicted to have a well-defined energetically preferred position to adsorb on top of a BP surface, that is, centered in the middle of the two top atoms along the armchair direction, with its longitudinal axis aligned along the zig-zag direction [29]. Based on this preferential positioning and considering steric constraints one can anticipate that there are two equally favorable possibilities for the arrangement of a monolayer of TCNQ molecules on top of BP, both having isosceles triangular symmetry with a base angle of either 36.4° or 65.7° (Supporting Information **Figure S1a** and **b**). Therefore, areas consisting of well-ordered arrangements of TCNQ molecules in both symmetries side-by-side can be principally achieved under controlled conditions, that is, in the limit of a low TCNQ pressure during sublimation and (comparably) short evaporation times[36]. In contrast, under high-pressure conditions and extended exposure times an oversaturation situation can be established. In this regime, there are two primary mechanisms working energetically against each other: (i) the formation of bi- and/or multilayers of TCNQ, and (ii) the formation of many, and initially not connected, (extended) clusters of TCNQ on the BP surface.

In the case when the formation of extended clusters is favored, a (local) de-wetting of the BP surface is principally possible, which then should lead to the formation of a fragmented layer of TCNQ molecules on the BP surface. Considering also that the preferential positioning of the first TCNQ molecules on the BP surface should remain partially preserved due to their strong binding energy, they could act as seeds for the layer fragments arranged in a geometry showing a specific periodicity.

The successful deposition of TCNQ on top of our BP flakes was confirmed by Raman spectroscopy and scanning probe microscopy. The Raman measurements showed the presence of characteristic peaks related to the TCNQ molecules [38,] besides the dominant modes associated with BP [39-41] (**Figure 1**). From the scanning probe microscopy the average thickness of the TCNQ layer was found to be 2 ± 1 nm (Supporting Information **Figure S2**). Also, the layer was found to be conformal to the initial surface of the BP.

We carried out a (time-dependent) observation of the formation of droplets under ambient conditions on BP samples with and without TCNQ. In **Figure 1c**, a comparison of AFM images is shown of an untreated, pristine BP flake (height h = 20 nm) and a BP flake with evaporated TCNQ on top (initial h = 13 nm), taken over several days. Clearly, on day 0 for neither of the two types of samples any particular droplet formation was observed. 24 hours later (day 1) irregularly appearing droplets were present on the pristine BP surface, whereas in the sample with deposited TCNQ no or only first signatures of a droplet formation were found. Here it has to be emphasized that TCNQ does not oxidize under our experimental conditions. On day 3 the droplets on the pristine sample have grown in size. On the sample with TCNQ discernible droplets are observable as well. However, the droplets of the latter are significantly smaller and appear to be more densely arranged compared to the pristine BP surface. 72 hours later (day 6), the droplets on the pristine BP surface have increased in size considerably and are randomly distributed as expected [21,22,30]. In contrast, the droplets on the sample with deposited TCNQ are significantly smaller and their density is higher. Moreover, their lateral size appears to be more regular than for the pristine BP surface. Another critical point regarding the droplets in the TCNQ treated BP sample is that the position of the droplets appears to have some degree of order. All these features point toward the possibility that a fragmented TCNQ layer has formed.

In **Figures 2a** and **2b** are shown the two-dimensional and three-dimensional representations, respectively, of the Fourier-transforms (FFT) of AFM images of a BP surface with and without TCNQ five days after exposure to ambient air. The results are qualitatively different: for the TCNQ treated sample a characteristic ring in the FFT amplitude is found. In contrast, for the pristine BP surface only a disk centered at around the zero wavevector is present. To extract the center wavevector of the ring, we averaged line-cuts of different quadrants (*cf.*

**Figure 2c**, right panels) of the Fourier transform and carried out the same analysis for the sample without TCNQ for comparison. The results are shown in the left panel of **Figure 2c**. For the pristine BP surface there is a continuous decay in FFT amplitude from zero μm$^{-1}$ towards higher wavevectors. This is the expected result for a system in which no periodicity is present. For the TCNQ treated BP sample, however, a clear maximum is found at about 6.2 ± 0.1 μm$^{-1}$ and no decay from zero wavevector. This implies that the system has a periodicity of about 161 ± 3 nm.

Considering that the presence of droplets on the surface is the result of the continued oxidation of BP, the existence of a periodicity corroborates the assumption that the evaporated TCNQ forms a fragmented layer-like structure. The circumstance that a ring in the FFT amplitude is observed implies that a by large isotropic periodicity is present which would be expected for TCNQ deposition conditions such as ours. Also, since the ring has a certain width, it reflects the circumstance that the TCNQ layer structure is not fragmented perfectly, that is, the fragments have some dispersion in size. However, these observations indicate that the layer fragmentation is similar to a dried up lake bed with fishing net like distribution of fragments, where the confined BP areas between fragments are exposed to the ambient. It is in these confined areas where the chemical reactions of the BP surface occur, and eventually droplets as fingerprint of the chemical reaction appear on top of the sample. In particular, the fishing-net like fragmentation is consistent with the two equally favorable arrangements of TCNQ on the BP surface (*cf.* **Figure S1**) in the monolayer limit. Consequently, the two arrangements of molecules can co-exist and act as seeds for extended cluster eventually leading to fragmented layers.

To gain further insight in the modification of the chemical reactivity within the confined areas between the fragments defined by the TCNQ network, the linear density *i.e.* number of droplets per unit-length, $n_d$, and the height of the droplets, $h_d$, were evaluated as function of time of exposure to air. In **Figure 3a** $n_d$ is shown for two pristine BP samples with height of 20 and 67 nm. We note that $n_d$ is here the average linear density taken over an ensemble of droplets measured on the surface.

In both cases, a very strong increase in $n_d$ is observed within the first 24 hours as expected for pristine BP [30-34], whereas afterwards a continuous decrease is found due to the coalescence of individual droplets into larger ones. Within the estimate error there is no or only a weak dependence of the oxidation process on the sample height (number of layers).

In **Figure 3b**, $n_d$ of two TCNQ treated samples with 13 and 40 nm height are shown. In contrast to the pristine BP surface, in both samples there is a rather gradual increase in the $n_d$ over four days until a maximum in $n_d$ is reached, followed by a rather sharp drop.

The difference in the oxidation dynamics can be further corroborated by comparing the average droplet height $h_d$ for all samples (**Figure 4**). For the pristine BP, a continuous pronounced increase in $h_d$ is found and shows indications of acceleration after day 4. In contrast, for the samples with TCNQ network only a rather moderate increase in $h_d$ is observed over a period of six-seven days.

The marked qualitative dissimilarity in the evolution of $n_d$ and $h_d$ with time for pristine and TCNQ exposed BP surfaces is a clear fingerprint of the difference in chemical reactivity within the confined areas between TCNQ layer fragments compared to an extended (quasi-infinite) BP surface.

In particular, it has to be emphasized, since the difference in the chemical reactivity as evidenced in **Figure 3** and **4** therefore cannot stem from a structural confinement effect such as may occur in ultra-small volumes defined by borders made of atoms in space or small particles of BP. Instead, another mechanism has to be present which is in accord with the preserved lattice of the two-dimensional surface.

Following this viewpoint, the TCNQ/BP surface interaction has to be recalled: A TCNQ molecule on top of a BP surface has the tendency for a strong charge transfer (~0.4 electron per molecule) [29] and the positive charge carriers introduced within the BP surface are strongly localized (flat-band) [35]. That is, there is no structural confinement by the TCNQ layer fragments, which could act on the electronic system, but only electrostatic confinement. Considering now that a chemical reaction is always associated with a modification of electrostatics, it is likely that confining two-dimensional surfaces electrostatically leads to a different dynamics of a chemical reaction. In our present case, based on the observations summarized in **Figure 3** and **4**, an inhibited oxidation of the BP surface within electrostatically confined areas is the result of such effects.

We demonstrated that by combining a naturally electrostatically pre-structured surface (as the one of BP) with the strong preferential positioning of molecules (TCNQ from a strongly over-saturated gas-phase), a large-scale periodically fragmented molecular layer can form spontaneously. Based on our time-dependent comparative study of the evolution of droplets on pristine and TCNQ-treated BP, it is suggested that the average TCNQ fragment lateral size is of the order of 100 nm. Interestingly, the most probable symmetry of a TCNQ monolayer on BP is in agreement with the apparent symmetrical arrangement of the fragments as reflected by the phosphoric and phosphorus acids/water droplet observed and deduced from the Fourier transforms of the experimental AFM images. This further supports that the initial arrangement, subsequent ordering and stacking of deposited molecules is associated with the specific surface electrostatics of BP and the preferential positioning of a TCNQ molecule on it. Our study of the number of droplets and their height as a function of time further revealed a marked qualitative difference in the dynamics of the oxidation process within the confined areas between TCNQ layer fragments as compared to a pristine BP surface. The differences observed are attributed to the local modification of the electrostatic conditions with respect to an extended (infinite) BP surface.

**Acknowledgements**

V.K. and A.H. thank the SFB 953 "Synthetic Carbon Allotropes" funded by the Deutsche Forschungsgemeinschaft (DFG) for support and the Cluster of Excellence „Engineering of Advanced Materials". A.H. acknowledges the European Research Council (ERC Advanced Grant 742145 B-PhosphoChem) for support. The research leading to these results was partially funded by the European Union Seventh Framework Programme under grant agreement No. 604391 Graphene Flagship. G.A. thanks the FAU for the Emerging Talents Initiative (ETI) grant #WS16-17_Nat_04, and support by the DFG and FLAG-Era (AB694/2-1).

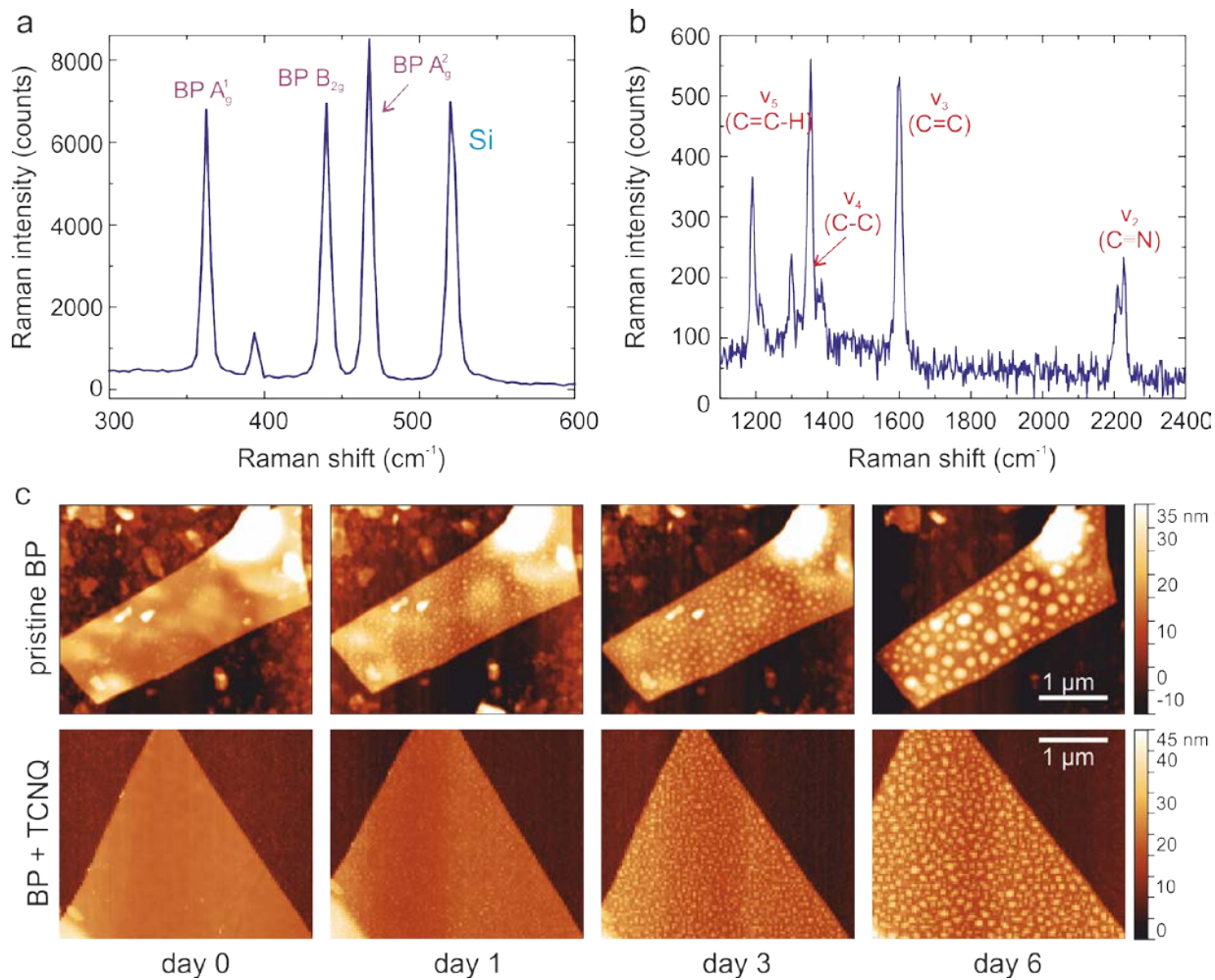

**Figure 1.** BP surfaces with and without TCNQ molecules. (a,b) Raman spectrum of BP with TCNQ on top showing (a) characteristic BP peaks in the range of 300 – 600 cm-1 [39,40] and (b) the presence of TCNQ-associated Raman modes in the 1100 – 2400 cm-1 range [38]. (c) Comparison of the time-elapsed oxidation process of pristine and TCNQ-treated BP surface. Top panel: AFM images of pristine BP exposed to air and monitored over six days (flake height h = 20 nm). The oxidation process ignites rapidly and leads to the formation of droplets distributed irregularly over the surface and increasing in size with time. Bottom panel: BP flake with deposited TCNQ on top (initial h = 13 nm). The formation of droplets is strongly inhibited. In particular, the droplets are on average significantly smaller in size at the same point in time compared to the pristine BP surface.

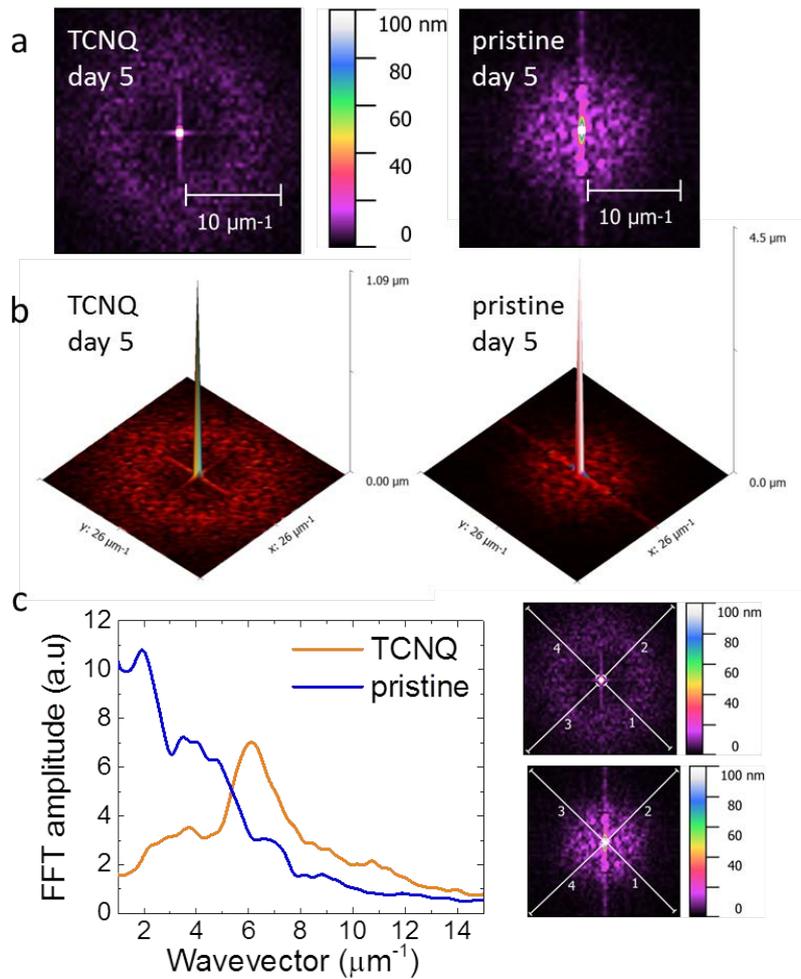

**Figure 2.** Fourier transforms (FFT) of AFM images of a TCNQ-treated and a pristine BP sample. a) Comparison of the FFT AFM images of TCNQ-treated (initial h = 13 nm) and pristine BP (h = 20 nm) after five days of exposure to ambient (same scale, same image sizes). While in the image of TCNQ/BP surface a well-defined ring feature is observed, for pristine BP only a disk-like structure around zero wavevector is found. b) 3D representations of the FFT in a). c) FFT amplitude averaged over line-scans (1 to 4, right panels) of four quadrants of the Fourier transformed AFM images. For pristine BP a continuous decay is observed, with no indication of a large-scale specific periodicity. In contrast, for the TCNQ-treated sample, a clear periodicity is observed centring at around 6 μm-1.

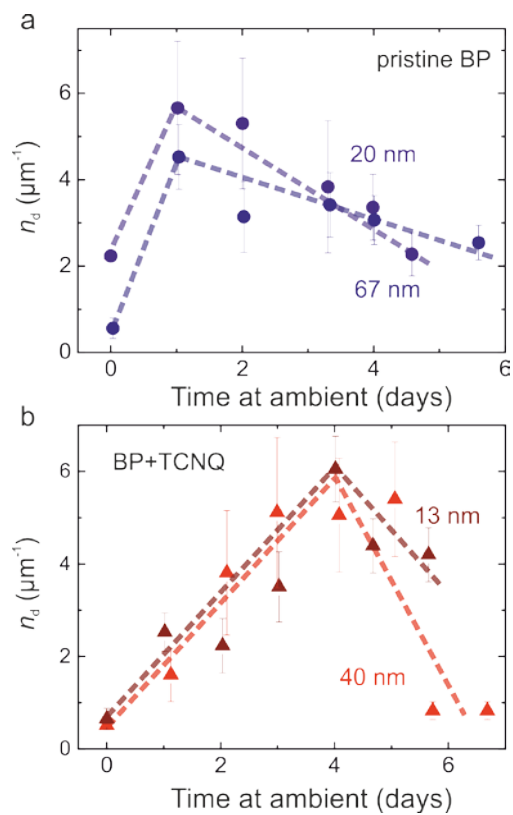

**Figure 3.** Evolution of average linear droplet density, $n_d$, with time. a) For pristine BP a very rapid increase in $n_d$ is observed within 24 h followed by a gradual decrease. In contrast, b) for a BP surface with TCNQ, a maximum in $n_d$ is only found after four days. Beyond that, indicators for a rather rapid decrease are present. Numbers in graphs: height of the flakes; lines: guide to the eye.

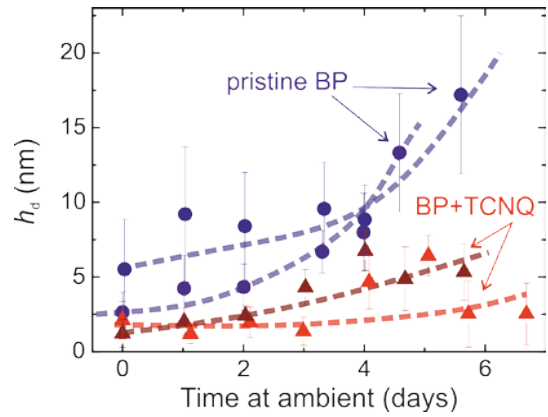

**Figure 4.** Evolution of average droplet height, $h_d$, with time. Symbols and color-coding following Figure 3. In the case of the pristine BP (blue) an accelerated increase of $h_d$ is observed beyond day 3 to 4. In contrast, the BP samples with TCNQ deposited on top (red) show a rather moderate increase over the whole duration of monitoring. Moreover, the $h_d$ magnitudes observed stay well below the ones for the pristine BP, specifically for beyond day 3. Lines: guide to the eye.

Supporting Information

**Effect of TCNQ layer cover on oxidation dynamics of black phosphorous Title**

*Laura Meingast, Maria Koleśnik-Gray, Martin Siebert, Gonzalo Abellán, Stefan Wild, Vinent Lloret, Udo Mundloch, Frank Hauke, Andreas Hirsch and Vojislav Krstić\**

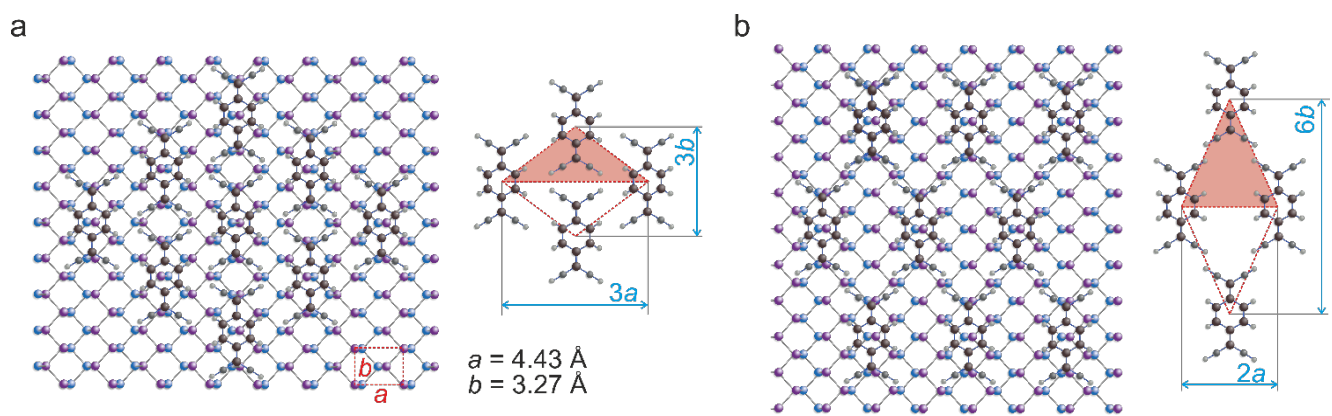

**Figure S1. Proposed arrangements of TCNQ molecules on BP (monolayer limit).** Considering the preferential positioning of a single TCNQ molecule along the zig-zag direction of the BP surface lattice and the actual size of a TCNQ molecule[1], the most probable arrangements of TCNQ molecules in the monolayer limit on the BP surface are isosceles triangular with base angle of **a)** 36.4° and **b)** 65.7°.

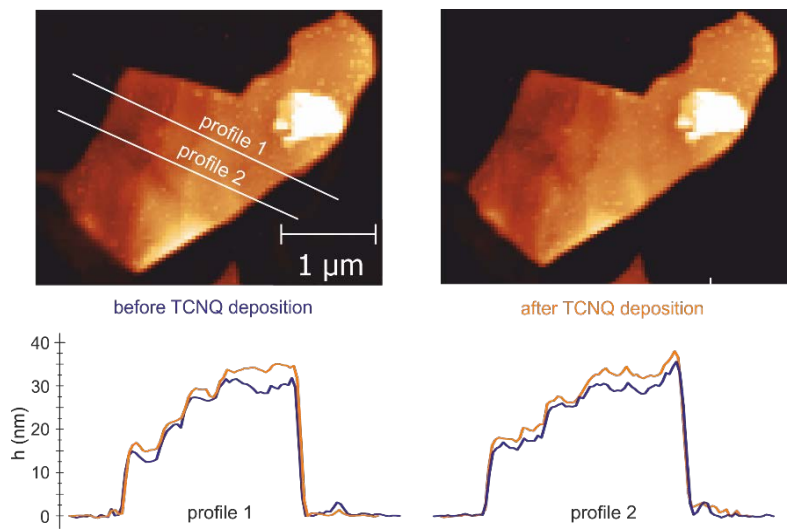

**Figure S2. AFM images and linescans of a BP flake before and after TCNQ.** Top- panel: On the left, the BP flake before the TCNQ deposition and linescan positions (profile 1 and 2) indicated. The AFM image on the right is the same flake with TCNQ layer. Bottom-panel: line-scans of the sample in the top-panel. The TCNQ layer is conformal to the surface (orange line after TCNQ deposition).

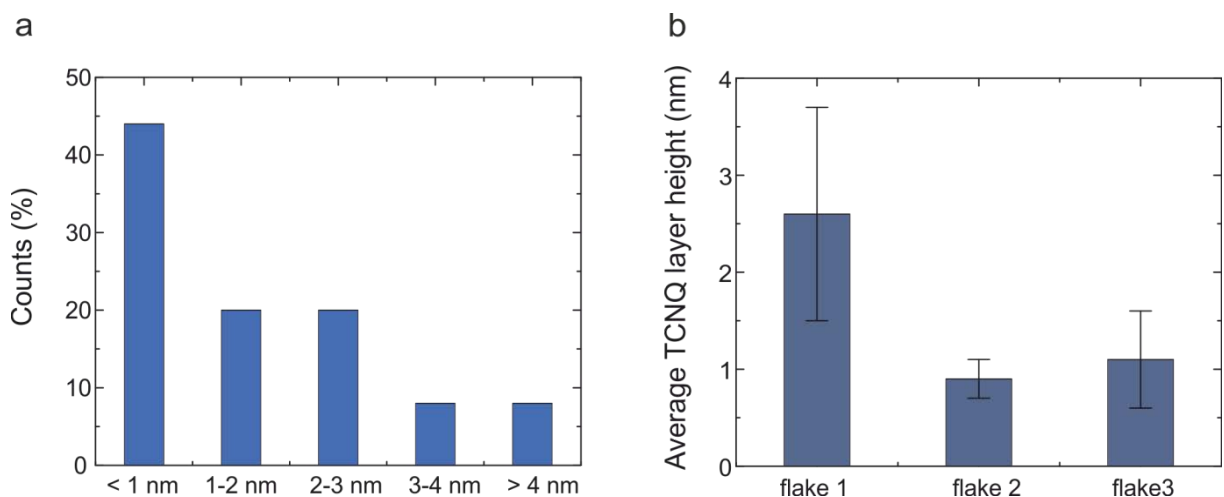

**Figure S3. Measured thicknesses of the TCNQ layer on BP surfaces. a)** Histogram showing the occurrence of the different heights measured by AFM on three BP samples with TCNQ deposited. **b)** Average height of the TCNQ layer as found in each of the samples. On average the TCNQ layer thickness is therefore $2 \pm 1$ nm.

Supporting References: